\documentclass[twocolumn,english,10pt,aps,prd,a4paper,preprintnumbers,floatfix,nofootinbib,showpacs,superscriptaddress, notitlepage]{revtex4-1} 
\pdfoutput=1

\usepackage{graphicx}
\usepackage{amsmath}
\usepackage{amssymb}
\usepackage{amsfonts}
\usepackage{physics}
\usepackage[dvipsnames]{xcolor}
\usepackage[linktoc=none]{hyperref}
\usepackage{braket}
\usepackage{tabstackengine}
\stackMath
\usepackage[mathscr]{euscript}

\hypersetup{colorlinks,linkcolor={blue},citecolor={teal},urlcolor={violet}}

\newcommand{\INFN}{INFN - Sezione di Napoli, Complesso Univ. Monte S. Angelo, I-80126 Napoli, Italy}
\newcommand{\UNINA}{Dipartimento di Fisica ``Ettore Pancini'', Università degli studi di Napoli ``Federico II'', Complesso Univ. Monte S. Angelo, I-80126 Napoli, Italy}
\newcommand{\SSM}{Scuola Superiore Meridionale, Università degli studi di Napoli ``Federico II'', Largo San Marcellino 10, 80138 Napoli, Italy}

\newcommand{\MPBH}{{M}_{\rm PBH}}

\newcommand{\TPBH}{\mathcal{T}_{\rm PBH}}

\newcommand{\GammaNth}{\Gamma_{N_1}^{\rm th.}}

\newcommand{\YB}{{Y}_{\rm B}}

\newcommand{\CED}{\mathcal{S}} 
\newcommand{\Sr}{\mathcal{S}_{\beta^\prime}/\mathcal{S}_{\beta^\prime = 0}}

\newcommand{\YNth}{\mathcal{N}_{N_1}}
\newcommand{\Yeq}{\mathcal{N}_{N_1}^{\rm eq.}}
\newcommand{\Yeqlep}{\mathcal{N}_{\ell}^{\rm eq.}}
\newcommand{\YBL}{\mathcal{N}_{\rm B-L}}

\newcommand{\YBm}{\tilde{Y}_{\rm B}}
\newcommand{\varrhopbh}{\varrho_{\rm PBH}}
\newcommand{\varrhorad}{\varrho_{\rm rad}}

\newcommand{\rhopbh}{\rho_{\rm PBH}}

\newcommand{\rhorad}{\rho_{\rm rad}}

\newcommand{\Tform}{T_{\rm in}}

\newcommand{\Mpl}{{M}_{\rm Pl}}

\newcommand{\rhocr}{\rho_{\rm cr.}}

\begin{document}

\title{Limits on light primordial black holes from high-scale leptogenesis}

\author{Roberta Calabrese}
\email{rcalabrese@na.infn.it}
\affiliation{\UNINA}
\affiliation{\INFN}

\author{Marco Chianese}
\email{marco.chianese@unina.it}
\affiliation{\UNINA}
\affiliation{\INFN}

\author{Jacob Gunn}
\email{jacobwilliam.gunn@unina.it}
\affiliation{\UNINA}
\affiliation{\INFN}

\author{Gennaro Miele}
\email{miele@na.infn.it}
\affiliation{\UNINA}
\affiliation{\INFN}
\affiliation{\SSM}

\author{Stefano Morisi}
\email{smorisi@na.infn.it}
\affiliation{\UNINA}
\affiliation{\INFN}

\author{Ninetta Saviano}
\email{nsaviano@na.infn.it}
\affiliation{\INFN}
\affiliation{\SSM}

\begin{abstract}
We investigate the role that the evaporation of light primordial black holes may have played in the production of the baryon asymmetry of the Universe through high-scale leptogenesis. In particular, for masses of primordial black holes in the range [$10^6$-$10^9$]~g, we find a dilution of thermally generated lepton asymmetry via entropy injection in the primordial plasma after the sphaleron freeze-out. As a consequence, we can put strong constraints on the primordial black hole  parameters, showing the mutual exclusion limits between primordial black holes and high-scale leptogenesis. Remarkably, we point out an interplay between the upper bound on the initial abundance of primordial black holes and the active neutrino mass scale.
\end{abstract}
\maketitle

\section{Introduction}
Primordial Black Holes (PBHs) are hypothetical black holes formed in the earliest times of the Universe, as a result of inflationary scenarios or physics beyond the Standard Model (BSM), and which are for this reason named primordial~\cite{Hawking:1974rv,Carr:1976zz,Khlopov:2008qy,Green:2020jor}. Since they are assumed not to be of stellar origin, PBHs can have very large mass ($10^{50}$\,g) but also very small mass, of the order of the Planck mass ($10^{-5}$\,g)~\cite{Carr:2009jm}. The awareness that PBHs might be very small, convinced Hawking to investigate their quantum properties. As a mixture of quantum field theory and general relativity, PBHs emit Hawking radiation in the form of a gray-body spectrum with a temperature $\TPBH$ inversely proportional to their mass. In other words, a black hole loses mass constantly by democratically emitting all the particles that are lighter than its temperature~\cite{Hawking:1975vcx, Hawking:1974rv}.
PBHs with masses above (of order of) $10^{15}$\,g  have not evaporated yet (are evaporating today) and have some chance to be detected, while lighter PBHs ($ \MPBH < 10^{15}$\,g) should have been evaporated by now giving potentially access to the physics of the early Universe~\cite{Carr:2020gox,Carr:2021bzv,Green:2020jor}, such as the physics of inflation and the primordial cosmological perturbations, baryogenesis and leptogenesis, the Big Bang Nucleosynthesis (BBN) physics, the physics of the cosmic microwave background, dark matter , and primordial gravitational waves     ~\cite{Byrnes:2021jka,Papanikolaou:2022did,Ozsoy:2023ryl,Ballesteros:2017fsr,Haque:2023rfp,Franciolini:2018vbk,Dolgov:2000ht,Chaudhuri:2017icn,Smyth:2021lkn,Baumann:2007yr,Hook:2014mla,Fujita:2014hha,Hamada:2016jnq,Hooper:2020otu,DeLuca:2021oer,Datta:2020bht,Wu:2021gtd,Perez-Gonzalez:2020vnz,Bernal:2022pue,Chen:2019etb,Morrison:2018xla,Chapline:1975ojl,LuisBernal:2017fmf,Keith:2020jww,Miyama:1978mp,Poulin:2016anj,Ali-Haimoud:2016mbv,Piga:2022ysp,Poulin:2017bwe,DeLuca:2021hcf,Bartolo:2018evs,Cheek:2021odj,Bird:2016dcv,DeRomeri:2021xgy,Dasgupta:2019cae,Mena:2019nhm,Lehmann:2019zgt,Biagetti:2021eep,Villanueva-Domingo:2021spv,Smyth:2019whb,Garcia-Bellido:2017aan,Kuhnel:2017bvu,Sasaki:2018dmp,Ireland:2023avg}. In particular, as initially investigated by~\cite{Hawking:1974rv,Carr:1976zz}, and later analysed in detail by~\cite{Baumann:2007yr},
the observed matter-antimatter asymmetry in the Universe could be generated by the PBH evaporation mechanism, given the emission of new heavy particles which can then decay violating CP and baryon number. The role of PBH evaporation on baryogenesis has recently raised new interest  in different contexts~\cite{Hook:2014mla,Smyth:2021lkn,Wu:2021gtd,Fujita:2014hha,Hamada:2016jnq,Morrison:2018xla,Hooper:2020otu,Datta:2020bht,DeLuca:2021oer, Gehrman:2022imk, Barman:2021ost, Barman:2022pdo, Barman:2022gjo}. 
Soon after the discovery of neutrino masses, the
type-I seesaw mechanism has acquired a growing popularity and with that also the simple and elegant mechanism of baryogenesis via leptogenesis from the first proposal by Fukugita and Yanagida~\cite{fukugita1986barygenesis}, where, instead of creating a baryon asymmetry directly from processes which violate the baryon number (B), a lepton asymmetry (L) should be generated from lepton number violating interactions before the electroweak phase transition (EWPT)~\cite{Davidson:2008bu, Fong:2012buy,Buchmuller:2005eh,DiBari:2012fz}.
Consequently, the lepton asymmetry is converted to a net excess of baryons via the non-perturbative (B + L)-violating sphaleron transitions~\cite{Harvey:1990qw,Kuzmin:1985mm,Khlebnikov:1988sr}.

The presence of primordial black holes in the early Universe could have impacted the process of leptogenesis mainly in two ways depending on the mass and so on the temperature of the PBHs, namely on the period relative to the leptogenesis era: {\it i)} through the production of heavy right-handed neutrinos, via the Hawking evaporation, in addition to the production from thermal plasma, {\it ii)} injection of entropy in the plasma, due to a production of significant population of photons by evaporation, effectively reducing the baryon-to-photon ratio.
The black hole induced leptogenesis scenarios have been well studied in literature~\cite{Perez-Gonzalez:2020vnz, JyotiDas:2021shi,Bernal:2022pue} where the several effects and their mutual impact are taken into  account for different choices of masses of right-handed neutrinos and PBHs. Among the different scenarios of leptogenesis, a special case is represented by high-scale leptogenesis, based on type-I seesaw mechanism, having hierarchical heavy Majorana right-handed neutrinos, $M_1\ll M_2\ll M_3$ with seesaw scale $M \ge 10^{ 12}$ GeV. Differently from what has been done in literature \cite{Perez-Gonzalez:2020vnz,Bernal:2022pue}, we focus only  on the second aspect induced by PBHs, namely the injection of entropy occurring after the sphaleron freeze-out. Indeed, choosing the range of PBH masses [$10^6$-$10^9$]~g, the production by evaporation of right-handed neutrinos is suppressed with a consequent negligible effect on the final asymmetry while a significant population of photons is produced  associated with an important entropy production in the primordial plasma and leading to a reduction of the baryon-to- photon ratio. We emphasize that the absence of the interplay of the  the aforementioned aspects, allows  us to scan the free parameters of the model as having the highest possible asymmetry permitted by high-scale leptogenesis and so to derive the entropy injection from PBH evaporation and to quantify the dilution of baryon asymmetry. In this way, we can put strong constraints on the PBH parameters, showing the mutual exclusion limits between PBHs and high-scale leptogenesis in the  plane of mass and abundance of PBHs. Indeed, we prove that, if the entropy injection is large enough, even the maximum value of the asymmetry permitted by the model of leptogenesis, would not reproduce the observed asymmetry of the Universe, ruling out leptogenesis as the baryogenesis mechanism. We also point out that this tension  between leptogenesis and light PBHs interestingly depends on the mass scale of active neutrinos with the higher the neutrino mass scale the stronger the PBH constraints. 

The structure of this work is as follows. In section~\ref{sec:HSL}, we describe the leptogenesis scheme used in our study and we present the Boltzmann equations which we solve for the baryon asymmetry. In section~\ref{sec:PBH_cosmology}, we outline how PBHs affect high-scale leptogenesis by altering the cosmological evolution. Section~\ref{sec:res} presents our results and some discussions, while section~\ref{sec:concl} contains our conclusions.

\section{High-scale leptogenesis model}\label{sec:HSL}

In this section, we outline the high-scale, thermal leptogenesis mechanism used throughout our analysis. Thermal leptogenesis is based on a minimal extension of the Standard Model (SM) comprising at least two singlet Majorana neutrinos which couple to the SM lepton and Higgs doublets. An attractive feature of this minimal model is the simultaneous explanation of the neutrino masses experimentally measured~\cite{Capozzi:2021fjo, Esteban:2020cvm, deSalas:2020pgw} and the baryon asymmetry of the Universe.

\subsection*{Type-I seesaw}
The small masses of the light, active neutrinos can be naturally generated in a type-I seesaw mechanism with the Lagrangian terms given by
\begin{equation}
    \mathcal{L} \supset - Y_{\alpha i} \overline{L}_\alpha  \tilde{\phi} N_i  -\frac{1}{2} \overline{N^c}_i \hat{M}_{ij} N_j + {\rm h.c.}\,,
\label{Lagrangian}
\end{equation}
where $\tilde{\phi} = i \sigma_2\phi^*$, $\phi$ is the Higgs doublet, $L_\alpha = (\nu_\alpha,\ell_\alpha)$ are the lepton doublets where $\alpha=e,\mu,\tau $, and $N_i$, $i=1,2,3$ are the Majorana right-handed neutrinos. The active neutrino mass matrix is then given by the type-I seesaw relation \cite{Yanagida:1980xy,MohapatraRabindraSenjanovi}
\begin{equation}\label{TI}
m_\nu\simeq -v_{\rm EW}^2\, Y\cdot\frac{1}{\hat{M}}\cdot Y^T \,.
\end{equation}
where $v_{\rm EW}=174$\,GeV is the Higgs vacuum expectation value, $\hat{M}$ is the mass matrix of the right handed neutrinos, and $Y$ is the Yukawa matrix appearing in Eq.~\eqref{Lagrangian}. One can always choose to work in the mass eigenbasis of the right-handed neutrinos so that $\hat{M} = \mathrm{diag}(M_1, M_2, M_3)$, while the $3\times 3$ matrix $m_\nu$ is diagonalised by the lepton mixing matrix $U_{\rm PMNS}$~\cite{pontecorvo1957mesonium, pontecorvo1957inverse, Maki:1962mu, Pontecorvo:1967fh, Gribov:1968kq} which is parameterised by three mixing angles, plus one Dirac and two Majorana phases. The relation~\eqref{TI} can be inverted to express the un-physical parameters $Y$ in terms of the physical ones~\cite{Casas:2001sr}
\begin{equation}\label{Y}
Y = \frac{1}{v_{EW}}\sqrt{\hat{M}}\cdot R \cdot \sqrt{\hat{m}_{\nu}} \cdot U^{\dagger}_{PMNS} \,,
\end{equation}
where $\hat{m}_\nu = \mathrm{diag}(m_1, m_2,  m_3)$,is the diagonalised left-handed neutrino mass matrix, and $R$ is an arbitrary orthogonal matrix parameterised by three complex angles $z_{12},z_{13},z_{23}$. 
While the atmospheric and solar neutrino oscillation data provide quite precise measurements of the PNMS mixing angles and of the two square mass differences, namely $\Delta m_{\rm sol}^2$ and $\Delta m_{\rm atm }^2$, the absolute scale of the neutrino mass and the neutrino mass ordering (normal: $m_1<m_3$ or inverted: $m_1>m_3$) are still unknown. 
In the following, we consider only normal mass ordering so that the heaviest and the lightest active neutrino masses $m_h ,\,m_l$ are identified with $m_3 ,\,m_1$, respectively. Moreover, We employ a minimal model of thermal leptogenesis with a hierarchical mass spectrum for the right-handed neutrinos so that $M_1\ll M_{2,3}$.

\subsection*{CP asymmetry}
The CP violation in the lepton sector plays a crucial role in the production of the baryon asymmetry via leptogenesis. In particular, a non zero CP asymmetry can be generated in the out-of-equilibrium decays of the heavy right-handed Majorana neutrinos.
In the model under consideration, the dominant production of asymmetry is associated with the decay of $N_1$ into the SM leptons and Higgs, $N_1\to \ell_\alpha \, \phi$, as long as the decay mode is CP asymmetric when evaluated at the one-loop level~\cite{Davidson:2008bu}. The amount of CP violation in the decay of $N_1$ into lepton flavour $\alpha$ is given by
\begin{equation}
\epsilon_{\alpha\alpha}=\frac{\Gamma(N_1\to \ell_\alpha \, \phi)-\Gamma(N_1\to \overline{\ell}_\alpha\, \overline{\phi})}{\Gamma(N_1\to \ell_\alpha \,\phi)+\Gamma(N_1\to \overline{\ell}_\alpha\, \overline{\phi})}\,,
\end{equation}
where $\Gamma$ indicates the width of the decay mode. For $T \gtrsim 10^{12}$\,GeV the charged lepton Yukawa interactions are slower than the Hubble rate, and leptogenesis is blind to the flavours of the leptons \cite{Davidson:2008bu}. In our analysis, we consider $10^{10}  \le M_1/{\rm GeV} \le 10^{16}$, but we neglect the flavour effects since our key results concern only $M_1 \gg 10^{12}$\,GeV. Therefore, summing over $\alpha = e,\mu,\tau$, the total CP asymmetry is (in the limit $M_1\ll M_{2,3}$)~\cite{Strumia:2006qk}
\begin{equation}\label{eps}
\epsilon=\sum_\alpha \epsilon_{\alpha\alpha} = - \frac{3}{16\pi}\sum_{j=2,3} \frac{M_1}{M_j}\frac{\mathrm{Im}(Y Y^\dagger)_{1j}}{(Y Y^\dagger)_{11}} \,.
\end{equation}
Note that in Eq.~\eqref{eps} the leptonic mixing matrix, $U_{\rm PMNS}$, has dropped out, thus implying no connection between the low-scale Dirac CP violation and high-scale CP asymmetry. Furthermore, as is shown in Ref.~\cite{Hambye:2003rt}, by approximating $m_1 = m_2$ since $\Delta m_{12}^2 \ll \Delta m_{23}^2$, rotations between the eigenstates $\nu_1$ and $\nu_2$, described by $R_{12}(z_{12})$, have no physical effects and the only relevant angle in the $R$ matrix in Eq.~\eqref{Y} is $z_{13}=x+i\,y$ where $x,y$ are real parameters. This allows us to express the $R$ matrix as
\begin{equation}
    R = \begin{pmatrix}
     \cos z_{13} &  0 & \sin z_{13} \\
     0 & 1 & 0\\
     -\sin z_{13} & 0 & \cos z_{13} \\
    \end{pmatrix}\,.
\end{equation}
With this simplification, the CP asymmetry parameter $\epsilon$ can be expressed in terms of only four unknown parameters $\{x,y, m_h, M_1\}$\,\cite{Bernal:2022pue}
\begin{equation}
    |\epsilon| = \frac{3M_1}{16\pi v_{EW}^2}\frac{|\Delta m^2_{\rm atm}|}{m_h + m_l}\frac{|\sin(2x)\sinh(2y)|}{\cosh(2y) - f\cos(2x)}\,,
\end{equation}
where $f \equiv (m_h - m_l)/(m_h + m_l)$ with $m_h = m_3$ and $m_l = m_1 \approx m_2$. While $\epsilon$ controls the amount of CP violation associated with the decay of $N_1$ particles, their thermal production violates CP with the same magnitude and opposite sign. Therefore when there is a vanishing initial abundance of $N_1$, an initial ``anti-asymmetry" is produced in the leptons. After reaching the thermal distribution, the population of $N_1$ will then decay when the temperature of the Universe is $T=M_1$, as long as the decay rate at that time is faster compared to the Hubble parameter. Such a condition is equivalent to~\cite{Strumia:2006qk,Davidson:2008bu, Buchmuller:2004nz}
\begin{equation}\label{K}
    K\equiv \frac{\tilde{m}_1}{m^*} > 1,
\end{equation}
where
\begin{equation}\label{meff}
    m^* = \frac{8\pi v_{EW}^2}{M_1^2} H\left(T=M_1\right)\,,
\end{equation}
and
\begin{equation}
    \tilde{m}_i = \sum_\alpha \frac{|Y_{\alpha\,i}|^2v_{EW}^2}{M_i}\,,
\end{equation}
representing the expansion rate at $T=M_1$ and the decay rate, respectively. When the right-handed neutrinos decay out of equilibrium, the resulting asymmetry would exactly cancel against the initial anti-asymmetry, if it were not for the so-called ``washout" processes which are crucial to leptogenesis.

Washout processes include the inverse decay modes of $N_1$, {\it i.e.} $\ell \phi^\dagger \to N_1$ and $\bar{\ell} \phi \to N_1 $, which are Boltzmann suppressed at low temperatures and do not contribute to the asymmetry, but are important at $T\sim M_1$ since they deplete the initial anti-asymmetry. Also important are $2\leftrightarrow2$ scattering processes like $\ell \phi^\dagger \to \bar{\ell} \phi $ mediated by $N_1$ exchange that do not suffer the same Boltzmann suppression since the intermediate state $N_1$ can go off-shell. For $K > 1$, the $N_1$ population decays when these processes are effective, so a non-zero lepton asymmetry survives since they wash out the initial anti-asymmetry, precluding any cancellation. For this reason, the condition $K > 1$ defines the ``strong-washout" regime. In the opposite regime $K < 1$, the decays are slow compared to the Hubble rate at $T = M_1$ and the $N_1$ population decays at a later time when the washout processes are weak, leading to a cancellation with the initial anti-asymmetry. In the weak-washout regime it is generally required to introduce a further mechanism to produce an initial population of $N_1$ to avoid this~\cite{Davidson:2008bu}. It is also interesting to note that the neutrino masses suggested by experiments favour the strong-washout regime since the weak-washout regime only appears for $m_h$ very close to $\sqrt{\Delta m^2_{\rm atm}}$. In our analysis, we always work in the strong-washout regime, so we require $|y| > 0.14$. This also means that any contribution to the asymmetry from $N_{2,3}$ is suppressed due to being washed out at $T\ll M_1$. 
\begin{figure*}[t!]
    \centering
    \includegraphics[width = 0.3 \textwidth]{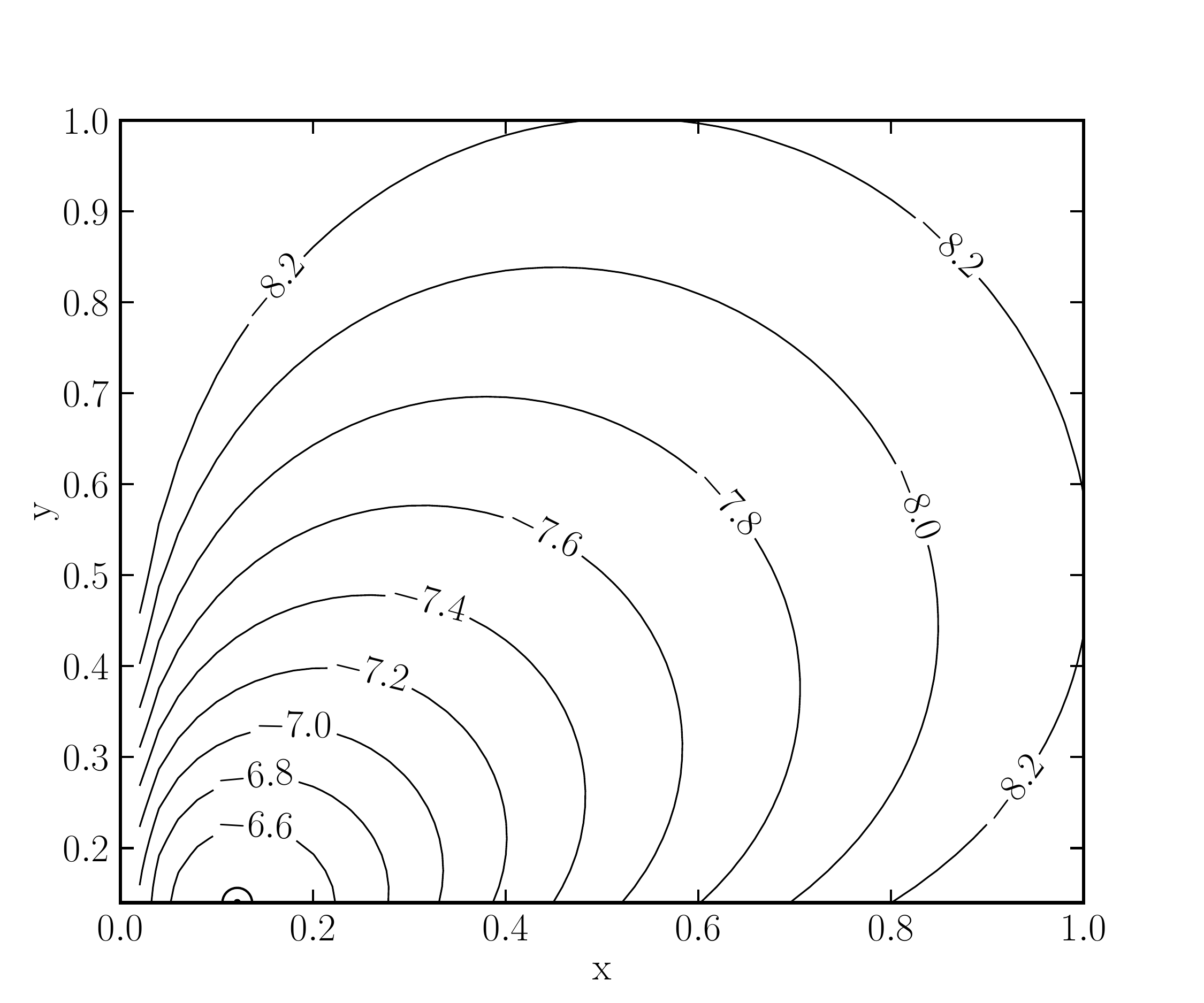}
    \includegraphics[width = 0.3 \textwidth]{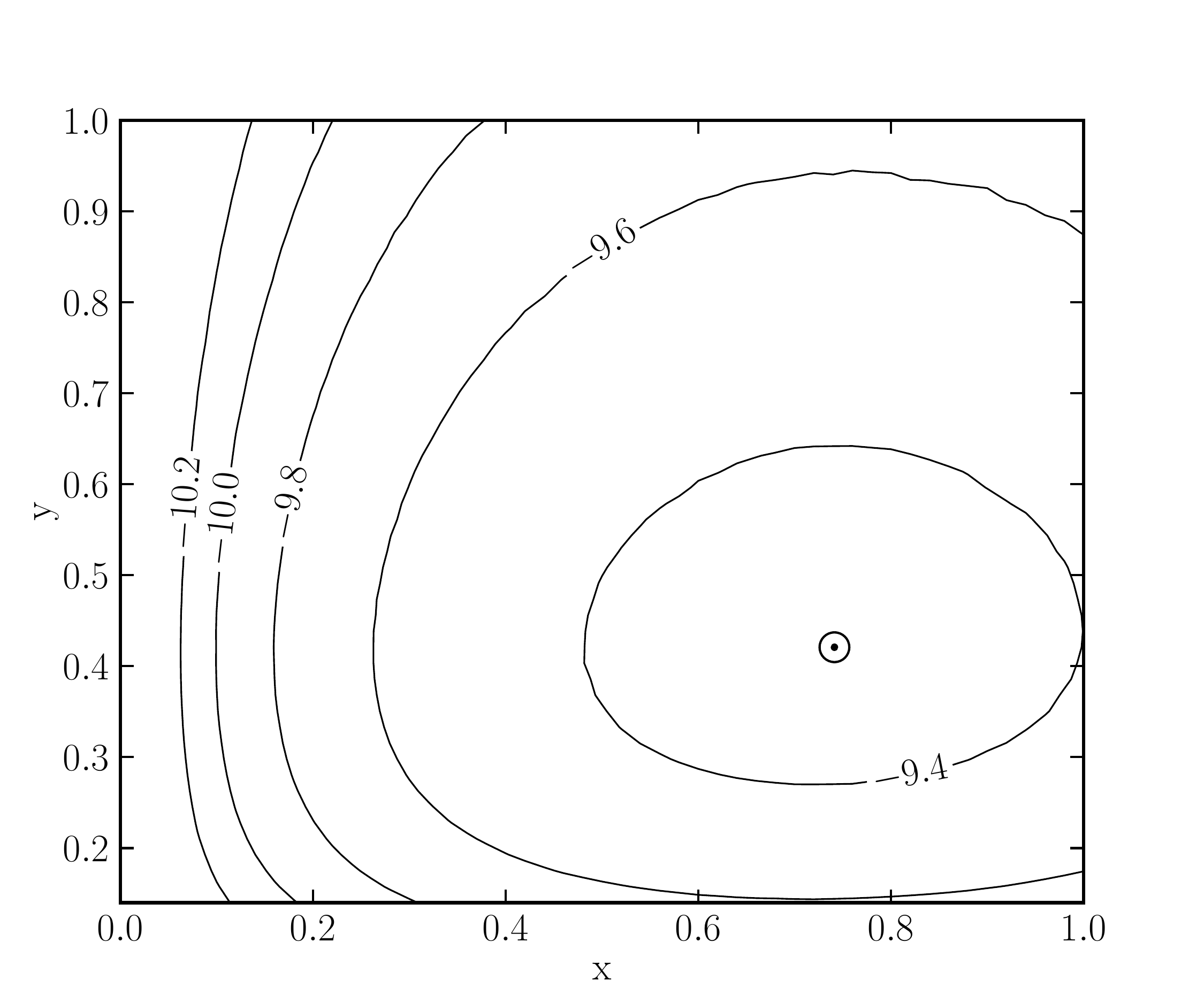}
    \includegraphics[width = 0.3 \textwidth]{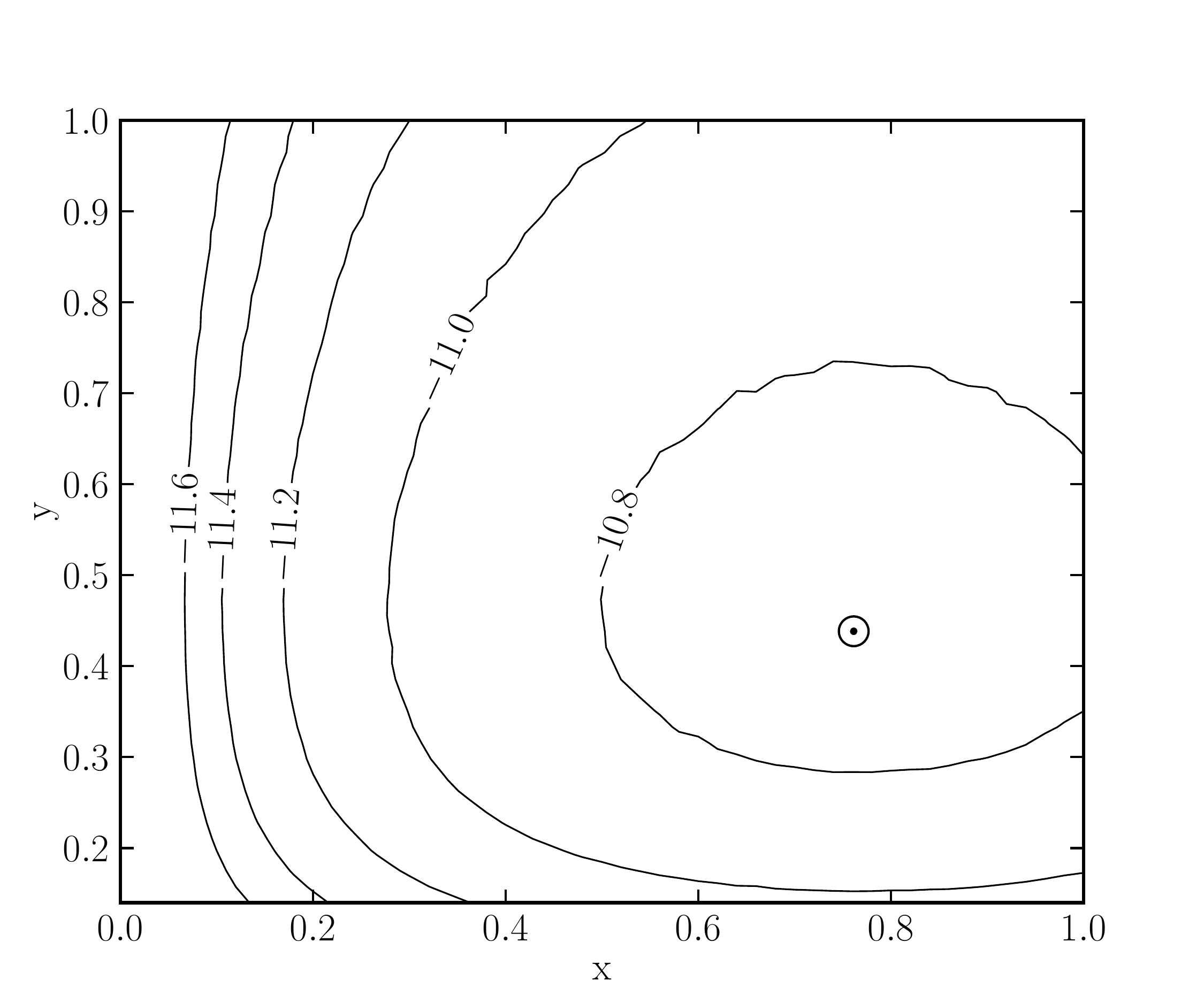}
    \caption{\label{xy}The final baryon asymmetry $\YB$ as a function of $x, \,y$ for $m_h = \sqrt{m_{\rm atm}^2} \approx 0.05$~eV (left panel), $m_h = 0.1$~eV (middle panel) and $m_h = 0.2$~eV (right panel), with $M_1 = 2.0 \times 10^{13}~{\rm GeV}$. The contours are for constant $\log_{10} \YB$ while the symbol $\odot$ indicates the point $(x,y)$ which maximises $\YB$ for the fixed values of $m_h$.}
\end{figure*}

\subsection*{Boltzmann equations}
To calculate the baryon asymmetry of the Universe, we solve numerically the Boltzmann equations describing the number density of $N_1$, and the $B-L$ asymmetry where $B$ and $L$ are the total baryon and lepton numbers of the Universe, respectively. It is convenient to track the evolution of the $B-L$ asymmetry since it is conserved by the sphaleron processes which partially convert the lepton asymmetry into a baryonic one.

In our analysis, we assume a vanishing initial abundance of $N_1$ and $B-L$ and we account for the following types of processes:
\begin{itemize}
    \item $1\to 2$ decays of $N_1$, $N_1 \to \ell \phi^\dagger$ and its CP conjugate process $N_1 \to \bar{\ell} \phi$. The total rates are proportional to the number density of $N_1$ and the thermally averaged decay width. Decays deplete the population of $N_1$ and contribute a source term to the $B-L$ asymmetry.
    \item $2 \to 1$ inverse decay modes like $\ell \phi^\dagger \to N_1$. The inverse decay rate is related to the decay rate by $\Gamma^{ID}_{N_1} = \Gamma_{N_1}n_{N_1}^{\rm eq}/n^{\rm eq}_{\ell}$ where $n_{N_1}^{\rm eq},n^{\rm eq}_{\ell}$ are the equilibrium number densities of $N_1$ and the leptons. These processes produce the $N_1$ population but only wash out the asymmetry.
    \item $2 \leftrightarrow 2$ scatterings mediated by $N_1$ exchange like $\ell \phi^\dagger \to \bar{\ell} \phi$, for which $\Delta L = 2$. These processes contribute to the washout and do not change the number density of $N_1$. Care should be taken when combining $2\leftrightarrow 2$ scatterings in the s-channel and the inverse decay rates to avoid double counting. A subtracted form of the amplitude should be used which only accounts for off-shell $N_1$ exchange, so these processes are not Boltzmann suppressed at low energies.
\end{itemize}
In principle one should also account for $2\leftrightarrow 2$ scatterings involving the gauge bosons and top quarks, but since these are order $(Y^2 Y_t^2)$ or higher in the couplings, where $Y_t$ is the top quark Yukawa coupling, we neglect them here. The effects of 3-body decay processes are also numerically small, $\sim 6\%$, and can be ignored~\cite{Nardi:2007jp}.

In standard $\Lambda$CDM cosmology, the resulting Boltzmann equations for the comoving number density of $N_1$ and $B$-$L$ are
\begin{eqnarray}
\frac{{\rm d}\YNth}{{\rm d}\alpha} &=&\ln(10)  \frac{\GammaNth}{H}(\Yeq-\YNth)\,, \label{eq:Nth} \\
\frac{{\rm d} \YBL} {{\rm d}\alpha} &=&   \left.\frac{\ln (10)}{H}\right[ \epsilon(\YNth-\Yeq)\GammaNth + \\ \nonumber
&&\qquad \left. + \left(\frac{1}{2} \frac{\Yeq}{\Yeqlep}\GammaNth + \gamma \frac{a^3}{\Yeqlep}\right)  \YBL\right]\,,\label{AsymBoltzmann}
\end{eqnarray}
where $\YNth = n_{N_1}a^3$, $\YBL = n_{\rm B-L}a^3$, $H = \Dot{a}/a$ is the Hubble rate and 
\begin{equation}
\alpha \equiv \log_{10}(a) \,,
\end{equation}
with $a$ being the scale factor. The quantity $\GammaNth$ is the thermally averaged, tree-level decay rate of $N_1$, given by 
\begin{equation}
    \GammaNth = \left<\frac{M_{1}}{E_{N_1}}\right>  \frac{(Y^\dagger Y)_{11}M_1}{8\pi}\,,
\end{equation}
where the thermal averaging factor $\left<M_{N_1}/E_{N_1}\right>$ is well approximated by the ratio $K_1(z)/K_2(z)$ for $K_n$ the n$^{\mathrm{th}}$ order modified Bessel functions of the second kind. In Eq.~\eqref{AsymBoltzmann}, $\gamma$ quantifies the contribution to the washout associated with the $\Delta L = 2$ scattering processes. In Ref.\,\cite{Bernal:2022pue} it is shown that 
\begin{equation}
    \gamma = \frac{3T^6}{4\pi^5 v_{EW}^4} \mathrm{Tr}[m_\nu^\dagger m_\nu]\,.
\end{equation}
Since the right-handed neutrino mass is much larger than the temperature of the electroweak phase transition $T_{\rm EWPT} \approx 160~{\rm GeV}$~\cite{DOnofrio:2015gop}, the lepton asymmetry stops evolving when the washout processes go out of equilibrium at temperatures far above the sphaleron freeze-out. The baryon asymmetry is then frozen-in when the sphaleron processes go out of equilibrium at $T = T^{\rm sph}$~\cite{DOnofrio:2014rug}, giving a yield of baryons today defined by
\begin{equation}\label{YB}
    |\YB| = \eta_{\rm sph} \frac{|\YBL(T=T^{\rm sph})|}{\mathcal{S}}\,.
\end{equation}
where $\mathcal{S}$ is the comoving entropy density of the Universe, given by $\mathcal{S} =2\pi^2g_*\left(aT\right)^3/45$ in a radiation dominated phase, where $g_*$ is the relativistic degrees of freedom, for $T\gtrsim 300$\,GeV, $g_* = 106.75$ in the SM~\cite{Kolb:1990vq}, and $\eta_{\rm sph}$ is the sphaleron efficiency factor and is taken to be 12/37~\cite{Harvey:1990qw}. The baryon asymmetry at present is described by the quantity $\eta_B=(n_B-n_{\overline{B}})/n_\gamma = (6.21\pm 0.16)\times 10^{-10}$ measured by the Planck collaboration~\cite{Planck:2018jri}, implying
\begin{equation}\label{YBexp}
 \YB^{\rm obs} = \frac{(n_B-n_{\overline{B}})}{s} = (8.75\pm0.23)\times 10^{-11}.     
\end{equation}
where $s$ is the entropy of the Universe.

\subsection*{The parameter space of high-scale leptogenesis}
All of the relevant quantities in Eq.s~\eqref{eq:Nth} and~\eqref{AsymBoltzmann}, and in particular today’s baryon asymmetry, can be now calculated in the four-dimensional parameter space $\{x,y, m_h, M_1\}$. We consider the following parameter ranges:
\begin{eqnarray}
0<x<\pi & \quad{\rm and}\quad & 0.14<y<\pi \label{range1}\\
0.05 < \frac{m_h}{{\rm eV}} < 0.8 & \quad{\rm and}\quad & 10^{10}< \frac{M_1}{{\rm GeV}} < 10^{16}\nonumber
\end{eqnarray}
The bound $y>0.14$ ensures the strong-washout regime in all the parameter space considered. The lower bound for the heaviest active neutrino mass $m_h$ comes from the neutrino oscillation data requiring $m_h > \sqrt{m_{\rm atm}^2}\approx 0.05~{\rm eV}$~\cite{Capozzi:2021fjo, Esteban:2020cvm, deSalas:2020pgw}, while the upper bound is set by the KATRIN experiment~\cite{KATRIN:2021uub}.

In Fig.~\ref{xy} we show the dependence of $\YB$ on the angles $x, \,y$ taking $M_1=2.0 \times 10^{13}~{\rm GeV}$, for three different active neutrino masses: $m_h = \sqrt{m_{\rm atm}^2} \approx 0.05$~eV (left panel), $m_h = 0.1$~eV (middle panel) and $m_h = 0.2$~eV (right panel). In the plots, the contours represent different values of $\log_{10} \YB$. Moreover, the symbol $\odot$ indicates the point $(x,y)$ which maximises $\YB$. For $m_h \gg \sqrt{m_{\rm atm}^2}$, the asymmetry $|\YB|$ is always maximised for $y = 0.44$ and $x = \pi/4$~\cite{Bernal:2022pue} (middle and right panels), while for smaller $m_h$ (left panel) the dependence on the angles $x,y$ is non-trivial requiring us to perform a detailed numerical scan. Between these two regimes, the point maximising $\YB$ is continuously translated from the position on the left panel to that on the right panel. As it will be detailed in the next section, the evaporation of primordial black holes leads to an entropy injection which dilutes the baryon asymmetry produced by leptogenesis. For this reason, we scan the parameter space to find the points which maximise the baryon asymmetry $\YB$. In particular, we choose to always fix the angles $x,y$ to maximise the final baryon asymmetry defining the quantity
\begin{equation}\label{YBm}
    \YBm(m_h,M_1)=\max_{x,y} Y_B (x,y,m_h,M_1)\,.
\end{equation}
\begin{figure*}[t!]
    \begin{minipage}[b]{0.5\linewidth}
    \centering
    \includegraphics[width=\linewidth]
    {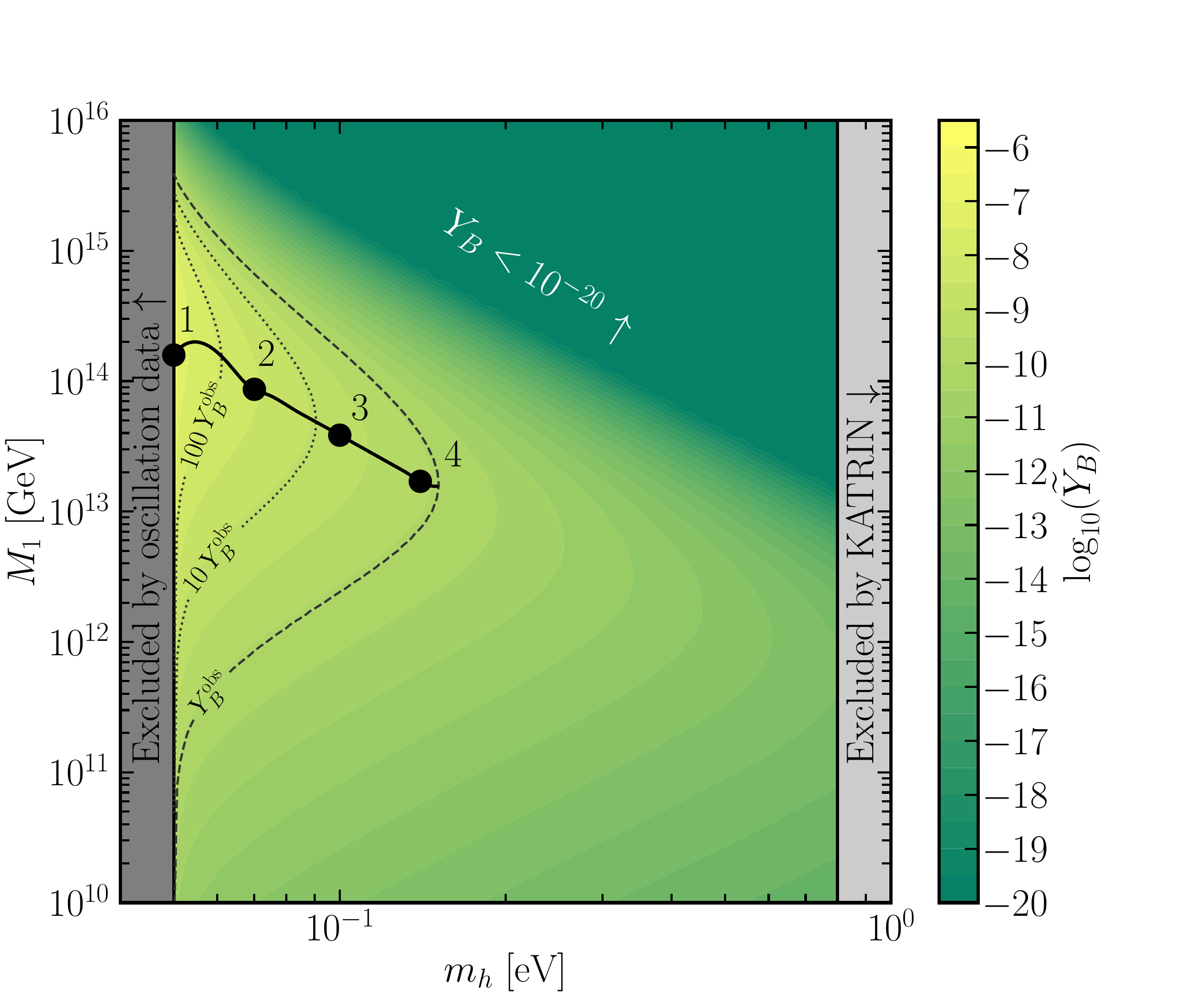}
    \end{minipage}%
    \hspace{0.05\linewidth}
    \begin{minipage}[b]{0.25\linewidth}
    \centering
    \begin{tabular}[b]{c|c|c|c}
        Bench. pt & $m_h~[{\rm eV}]$ & $M_1~[{\rm GeV}]$ & $\YBm$ \\ \hline
         {\bf 1} & $0.05$ & $1.5 \times 10^{14}$ & $1.5 \times 10^{-6} $ \\ \hline
         {\bf 2} & $0.07$ & $1.0 \times 10^{14}$ & $3.6 \times 10^{-9} $ \\ \hline
         {\bf 3} & $0.10$ & $4.0 \times 10^{13}$ & $5.5 \times 10^{-10}$ \\ \hline
         {\bf 4} & $0.14$ & $2.0 \times 10^{13}$ & $1.2 \times 10^{-10}$ \\ 
    \end{tabular}
    \vspace{80pt}
    \end{minipage}
    \hspace{0.05\linewidth}
    \caption{\label{fig:Ybm}The quantity $\YBm$ defined in Eq.~\eqref{YBm} as a function of $m_h$ and $M_1$. The dashed contour marks the observed baryon asymmetry of the Universe, while the dotted ones represent larger values as reported in the labels. The solid line shows the parameters that maximise the baryon asymmetry, with points from 1 to 4 highlighting the benchmark cases reported in the table on the right. The grey bands are excluded by experimental limits from oscillation data~\cite{Capozzi:2021fjo, Esteban:2020cvm, deSalas:2020pgw} (left) and KATRIN searches~\cite{KATRIN:2021uub} (right).}
\end{figure*}

In Fig.~\ref{fig:Ybm}, we report the maximum baryon asymmetry $\YBm$ in colour, as a function of the heaviest left and
lightest right handed neutrino mass scales. The dashed line represents the parameters for which $\YBm$ matches the observed value reported in Eq.~\eqref{YBexp}. It also delimits from the right the region in the parameter space where the high-scale thermal leptogenesis can correctly account for the baryon asymmetry of the Universe. Indeed, in that region the observed value $\YB^{\rm obs}$ can be achieved by simply tuning the angles $x,y$. In the plot, we also show different increasing values for the ratio $\YBm/\YB^{\rm obs}$ with dotted lines. Moreover, the solid line represents the relation between $m_h$ and $M_1$ that maximises the baryon asymmetry $\YB$. The numbered points highlight benchmark cases for different active neutrino mass $m_h$ that will be used as reference in later plots. For each value of $M_1$, the final baryon asymmetry increases for decreasing $m_h$. We find that the the maximal allowed value for the baryon asymmetry in the high-scale thermal leptogenesis scenario is
\begin{equation}\label{benchmark}
\YB^{\rm max} = 1.5\times 10^{-6}
\end{equation}
achieved for $x=1$, $y = 0.14$, $m_h = 0.05~{\rm eV}$ and $M_1 = 1.5 \times 10^{14}\,{\rm GeV}$.
For this choice of parameters, an entropy production of about four orders of magnitude at later times (e.g. from PBHs' evaporation) would be required to obtain the observed baryon asymmetry of the Universe (see Eq.~\eqref{YBexp}). A larger amount of entropy production would be in tension with the high-scale thermal leptogenesis giving $\YB < \YB^{\rm obs}$.

\section{Cosmology with Primordial Black Holes}\label{sec:PBH_cosmology}

We now describe the influence of PBH physics on the evolution of the Universe. In particular, PBH evaporation may cause a period of reheating in the early Universe, associated with significant entropy production\cite {Dolgov:2000ht}.\footnote{See \cite{Chen:2019etb} for the treatment of leptogenesis in another non-standard cosmological scenario.} Being inversely proportional to the entropy density, $\YB$ will be consequently diluted. If PBHs dominate the energy budget of the Universe, the Hubble parameter will also be enhanced according to a matter-dominated period. Proper treatment of leptogenesis in a PBH cosmology must account for each of these effects.

For the sake of concreteness, we consider neutral and non-rotating (Schwarzschild) PBHs with a monocromatic mass distribution.
We assume that a population of PBHs with mass $\MPBH$ forms from the collapse of density perturbations when the radiation-dominated Universe cools to 
\begin{equation}
    \Tform=\frac{1}{2}\,\left(\frac{5}{\pi^3 g_*}\right)^{1/4}\sqrt{\frac{3\gamma_{\rm PBH} \Mpl^3}{\MPBH}}\,.
    \label{eq:T_in}
\end{equation}
where $\gamma_{\rm PBH}\sim0.2$ is the gravitational collapse factor~\cite{Carr:1975qj, Carr:2016drx}, and $\Mpl = 1.22\times10^{19}$\,GeV is the Planck mass. The initial abundance of the PBHs is typically defined in terms of the parameter:
\begin{equation}
    \beta^\prime = \gamma^{1/2}_{\rm PBH}\left(\frac{g_*(\Tform)}{106.5}\right)^{-1/4}\left(\frac{h}{0.67}\right)^{-2}\left. \frac{\rhopbh}{\rhocr}\right|_{\rm in}\,\,.
    \label{eq:beta_prime}
\end{equation}
where $h = 0.67$\,\cite{Planck:2018jri} is related to the Hubble constant,
$\rhopbh$ and $\left.\rhocr \right|_{\rm in } = \frac{3\gamma_{\rm PBH}^2}{32\pi}\frac{\Mpl^6}{(\MPBH)^2}$ are the PBH and critical energy densities at the time of PBH formation, respectively. We note that $\rhocr \simeq \rhorad$ at $\Tform$. 

From the moment of formation until they evaporate completely, PBHs radiate away their mass as all the particles with mass below their Hawking temperature~\cite{Hawking:1975vcx, Hawking:1974rv}. For Schwarzschild black holes, the Hawking temperature is given by \cite{Hawking:1974rv}
\begin{equation}
    \TPBH \simeq 10^7 \left(\frac{10^{6} {\rm g}} {\MPBH} \right){\rm GeV}\,.
\label{PBH temperature}
\end{equation}
For PBH masses in the range from $10^6$ to $10^9$~g considered in this study, PBHs do not efficiently produce right-handed neutrinos heavier than $10^{10}~{\rm GeV}$ bur rather radiate all SM degrees of freedom in the form of a grey-body thermal spectrum defined by $\TPBH$. Moreover, we note that for $\MPBH < 10^9$ the complete PBH evaporation ends before the beginning of BBN which occurs at $T \simeq {\rm MeV}$.

The effect of PBHs on the evolution of the Universe is twofold. First, PBHs may dominate the energy density of the Universe, leading to an enhancement of the Hubble rate. Second, PBHs may reheat the thermal photon bath through the Hawking emission of relativistic SM particles having an energy higher than the photon temperature. These effects can be studied by solving the following set of coupled differential equations~\cite{Lunardini:2019zob, Perez-Gonzalez:2020vnz, Bernal:2022pue}
\begin{eqnarray}
\frac{{\rm d}\varrhorad}{{\rm d}\alpha} &=& - f_{\rm SM} 10^{\alpha} \frac{{\rm d}\ln \MPBH}{{\rm d}\alpha}  \varrhopbh\,, \label{eq:FE_rad}\nonumber\\
    \frac{{\rm d}\varrhopbh}{{\rm d}\alpha} &=& \frac{{\rm d}\ln \MPBH}{{\rm d}\alpha}\varrhopbh\,,
\label{eq:FE_PBH}\\
\frac{{\rm d}\ln \MPBH}{{\rm d}\alpha} &=& - \frac{\kappa(\MPBH)}{H \MPBH^3} \ln(10)\,,\label{dln}\nonumber\\
    H^2 &=& \frac{8\pi}{3\Mpl^2}\left(\frac{\varrhopbh}{a^3}+\frac{\varrhorad}{a^4} \right)\,\,,\label{H}\nonumber
\end{eqnarray}
where $\varrhopbh \equiv a^3 \rhopbh$ and $\varrhorad \equiv a^4 \rhorad$ are the comoving energy densities of PBHs and radiation respectively, $\kappa(\MPBH) = 416.3\,\Mpl^4/(30720\pi)$ is constant for $\MPBH\lesssim 10^{9}$\,g\,\cite{Datta:2020bht} and $f_{\rm SM}$ is the fraction of Hawking radiation composed of SM particles. In our framework, $f_{\rm SM} \simeq 1$ since PBHs produce a negligible amount of right-handed neutrinos. We observe that the standard cosmological evolution is recovered when $\varrhopbh = 0$. From the first differential equation, one can also get the following equation for the comoving entropy density:
\begin{equation}
    \frac{{\rm d}\CED}{{\rm d}\alpha} = -\frac{f_{\rm SM}}{T(\alpha)} \frac{{\rm d}\ln \MPBH}{{\rm d}\alpha} \varrhopbh\,.
    \label{eq:entropy}
\end{equation}
which vanishes as expected for $\varrhopbh \to 0$ , namely the comoving entropy density remains constant. The dependence of the photon temperature $T$ on $\alpha$ is approximately described by
\begin{equation}
    \frac{\dd T}{\dd\alpha} \simeq -T \left[\ln(10) - \frac14 f_{\rm SM} 10^\alpha \frac{\dd \ln \MPBH}{\dd\alpha}\frac{\varrhopbh}{\varrhorad}\right]\,.
    \label{eq:T_univ}
\end{equation} 
By solving such equations, one finds that after PBH evaporation $\mathcal{S}$ can be significantly larger than the standard entropy density in the absence of PBHs ($\beta^\prime = 0$). The amount of entropy injection can be simply quantified by the ratio $\Sr$.

\section{Results}\label{sec:res}
\begin{figure*}[t!]
    \centering
        \includegraphics[width = 0.49\textwidth]{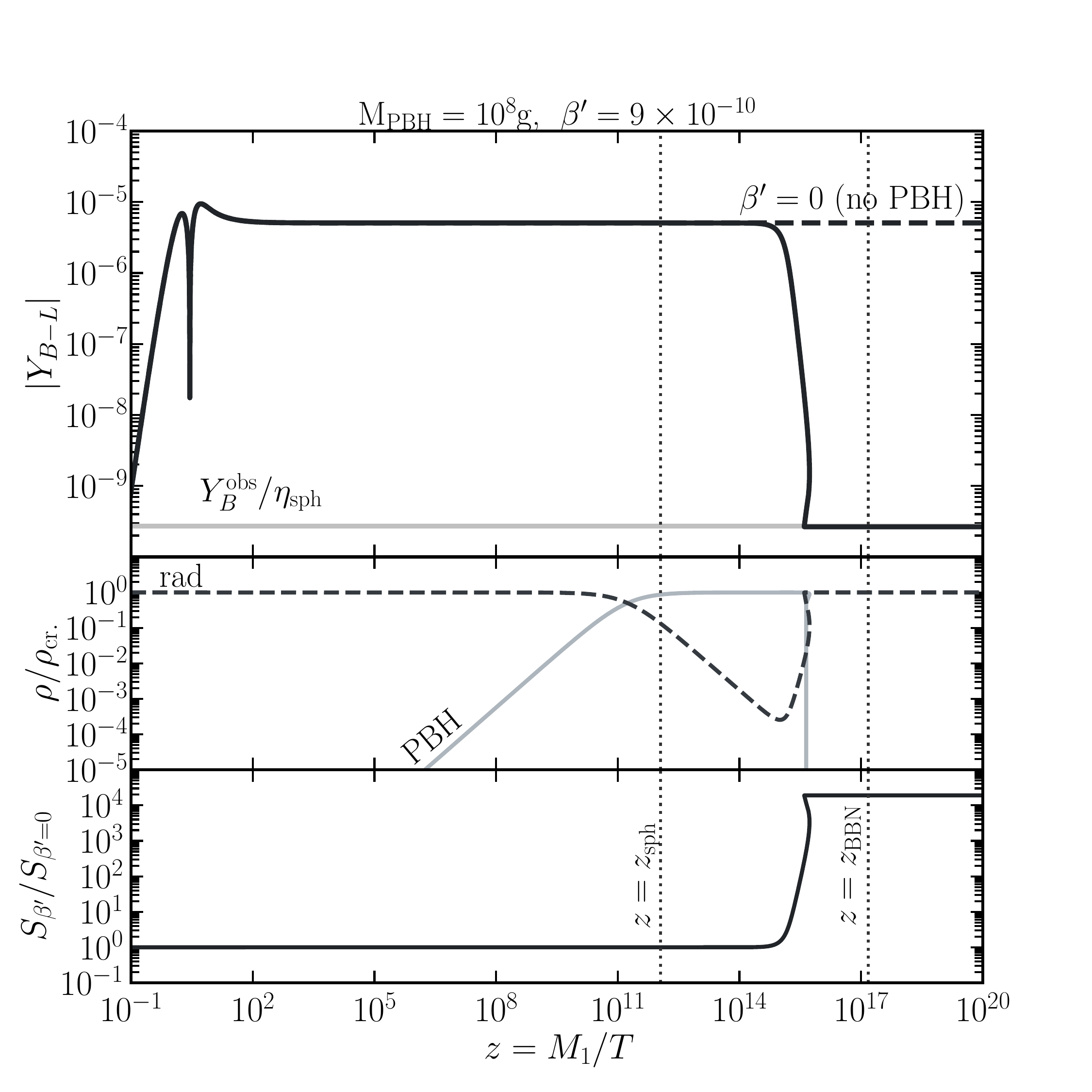}
        \includegraphics[width = 0.49\textwidth]{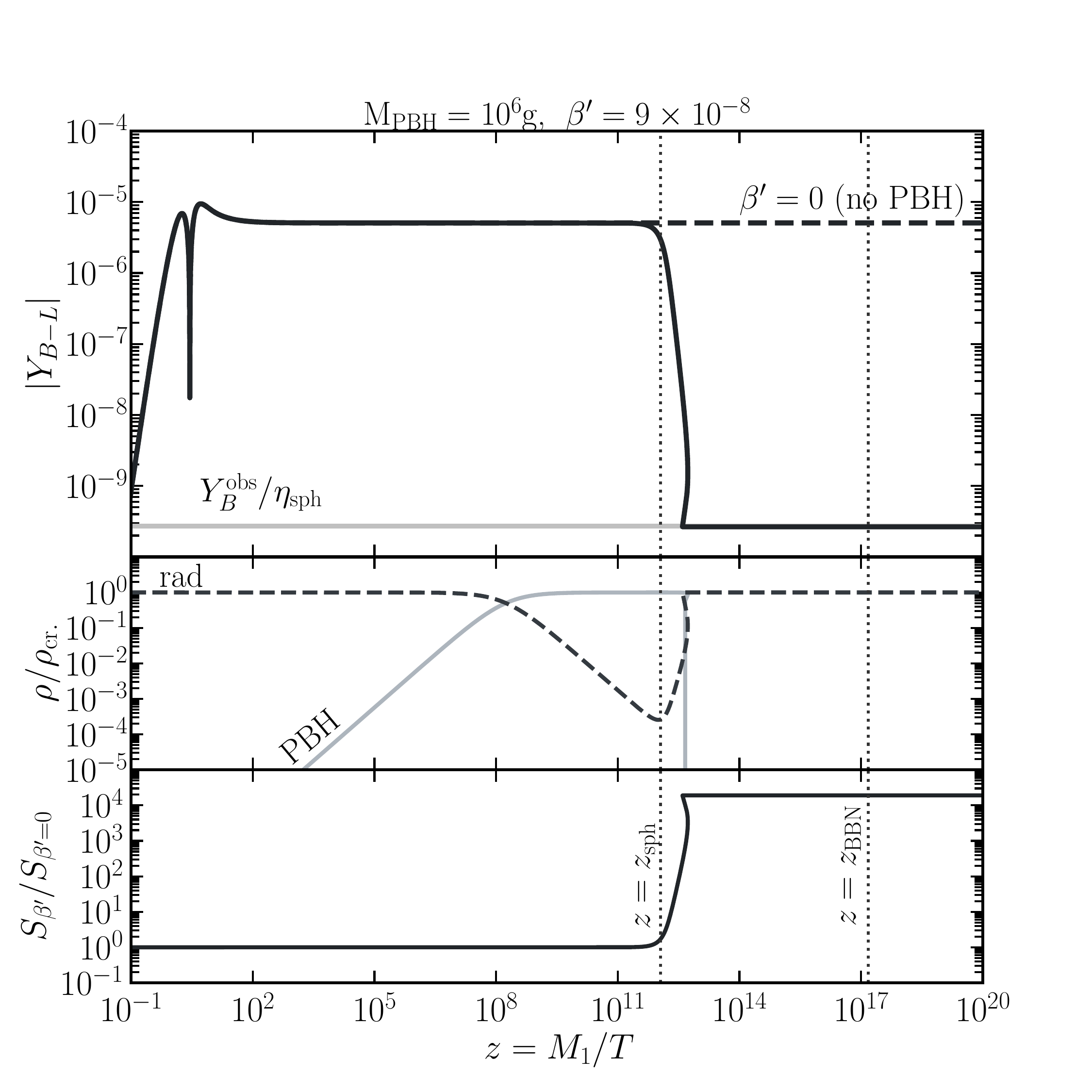}
       \caption{\label{fig:benchmark} The yield $Y_{\rm B-L}$ as a function of $z=M_1/T$ (top panels) with and without the presence of PBHs taking $\MPBH = 10^6~{\rm g}$ (left plot) and $\MPBH = 10^8~{\rm g}$ (right plot). The PBH initial abundance is fixed in order to obtain the observed baryon asymmetry (solid grey horizontal line) after PBH evaporation. The neutrino parameters are fixed according to the benchmark case 1 in Fig.~\ref{fig:Ybm}. The middle and bottom panels show the evolution of energy densities and the entropy ratio $\Sr$.}
\end{figure*}
\begin{figure*}[t!]
    \centering
        \includegraphics[width = 0.49\textwidth]{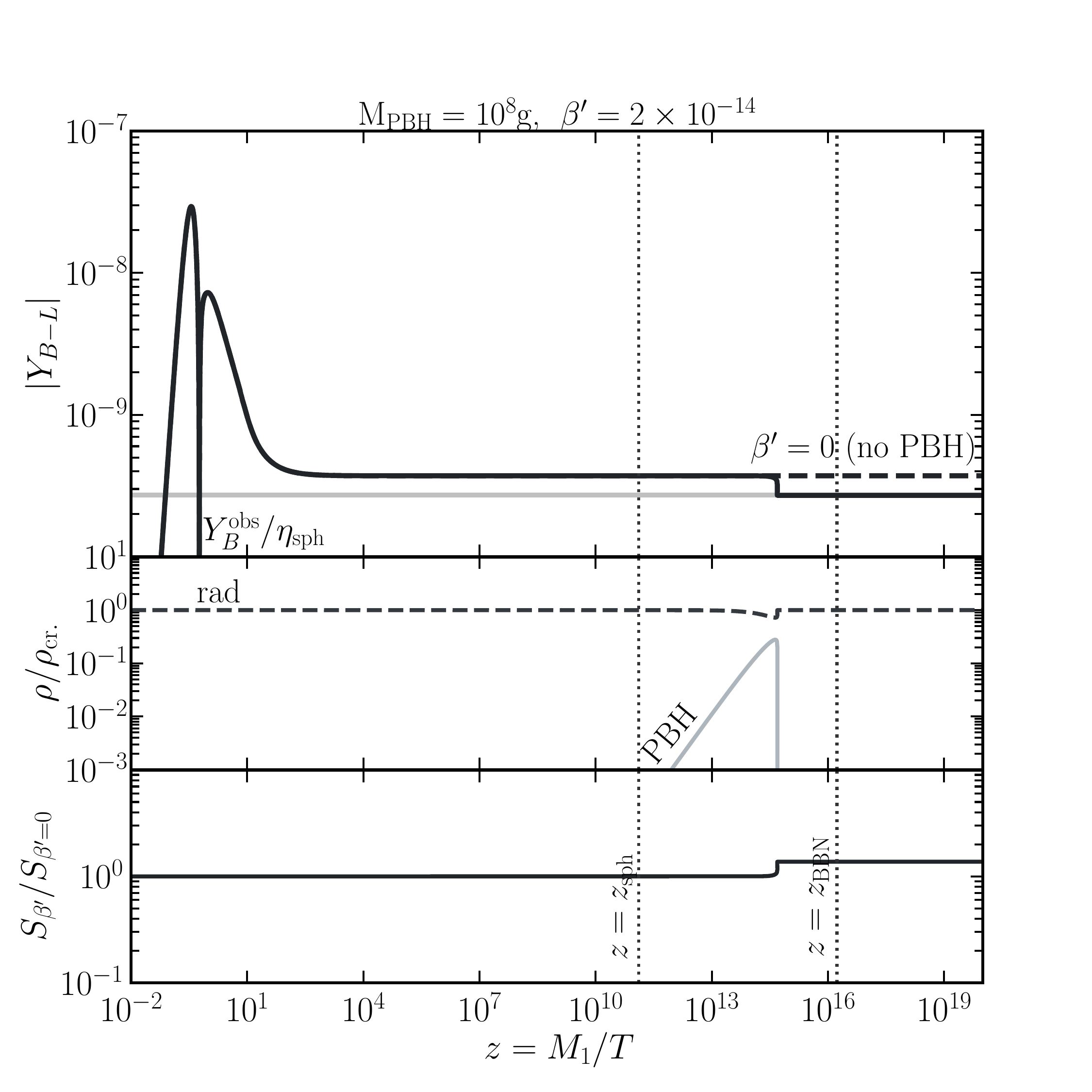}
        \includegraphics[width = 0.49\textwidth]{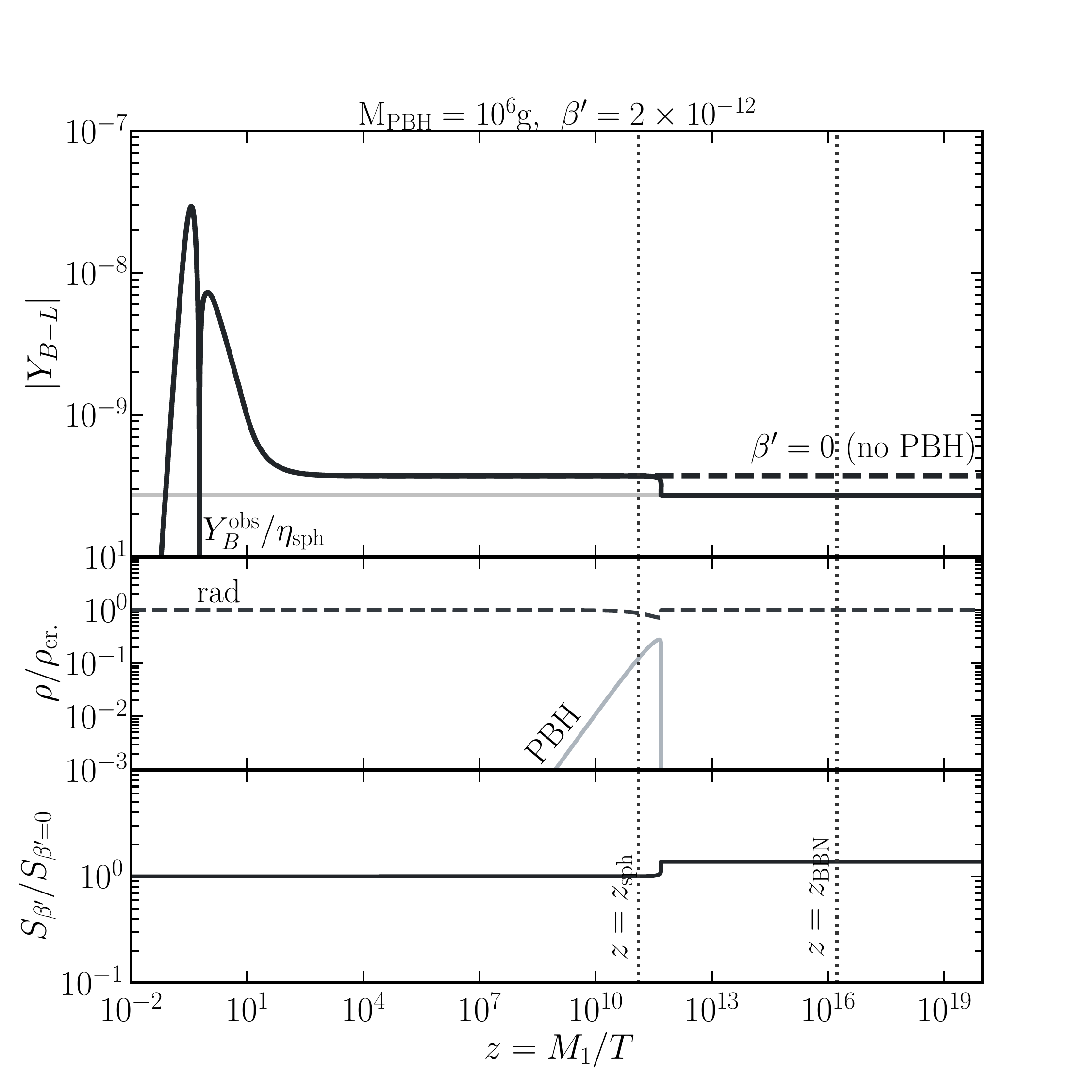}
       \caption{\label{fig:benchmark2} Same as in Fig.~\ref{fig:benchmark} but fixing the neutrino parameters according to the benchmark case 4 reported in Fig.~\ref{fig:Ybm}.}
\end{figure*}

Our model is described by six free parameters: four are needed to describe the leptogenesis mechanism $\{x,\,y,\, m_h,\,  M_1\}$, taken in the range in Eq.~\eqref{range1}, and two parameters that describe PBH physics $\{\MPBH,\, \beta^\prime\}$ in the range 
\begin{equation}
    10^6 < \MPBH/ {\rm g} < 10^9 \qquad {\rm and} \qquad 10^{-15}< \beta^\prime <0.1
    \label{range2}
\end{equation}
As already discussed, we perform a scan on the free parameters to identify the maximum $\YB$ attainable for fixed combinations of $m_h$ and $M_1$ (see Eq.~\eqref{YBm}). For each choice of point in our parameter space, we solve the two coupled equations~\eqref{eq:Nth} and~\eqref{AsymBoltzmann}, then by means of Eq.~\eqref{YB} we obtain $\YB$ today. The results are reported in Fig.s~\ref{xy} and~\ref{fig:Ybm}. Then, for each choice of $\MPBH$ and $\beta^\prime$, by solving Eq.s~\eqref{eq:FE_PBH} and~\eqref{eq:entropy}, we can derive the entropy injection from PBH complete evaporation and quantify the dilution of baryon asymmetry.

In Fig.s~\ref{fig:benchmark} and~\ref{fig:benchmark2}, we show how the presence of light PBHs alters the evolution of the $B$-$L$ yield (top panel) as a function of the auxiliary variable $z = M_1/T$. In the former figure, we fix the neutrino parameters according to the benchmark point 1 in Fig.~\ref{fig:Ybm}, while in the latter they correspond to the benchmark point 4. In the left (right) plots, the PBH mass is taken be to $10^6~{\rm g}$ ($10^8~{\rm g}$), with their initial abundance $\beta^\prime$ fixed to achieve the observed baryon asymmetry after evaporation. In the top panels of the plots, the dashed lines correspond to the standard evolution without PBHs for which the final asymmetry is higher than the observed one represented by the solid horizontal grey lines. We also report the evolution of the energy densities of radiation and PBHs (middle panels) and the ratio $\Sr$ (bottom panels). The left (right) plot refers to the benchmark point 1 (4) reported in the table in Fig.~\ref{fig:Ybm} with an initial PBH abundance of $\beta^\prime = 10^{-9}$ ($\beta^\prime = 3 \times 10^{-14}$). In Fig.~\ref{fig:benchmark} (benchmark point 1), PBHs rule a matter-dominated phase until they evaporate leading to a large entropy injection. In Fig.~\ref{fig:benchmark2} (benchmark point 4), even though PBHs never dominate the energy content of the Universe, their evaporation dilutes the final baryon asymmetry to match the observed value. In these benchmark cases, as well as in the whole PBH mass range considered (see Eq.~\eqref{range2}), the PBH evaporation occurs after the sphaleron processes freeze-out at $z = z_{\rm sph}$ (vertical dotted lines) and before the beginning of BBN at $T \simeq 1~{\rm MeV}$. Therefore, PBHs mainly modify the cosmological evolution only after the $B$-$L$ asymmetry has already been generated via leptogenesis. We note that the energy density of PBHs linearly depends on their initial abundance $\beta^\prime$. This implies that values for $\beta^\prime$ larger than the ones adopted in such benchmark cases would lead to $\YBm < \YB^{\rm obs}$ and consequently would be in tension with the minimal high-scale leptogenesis scenario.
\begin{figure}[t!]
    \centering
        \includegraphics[width =\linewidth]{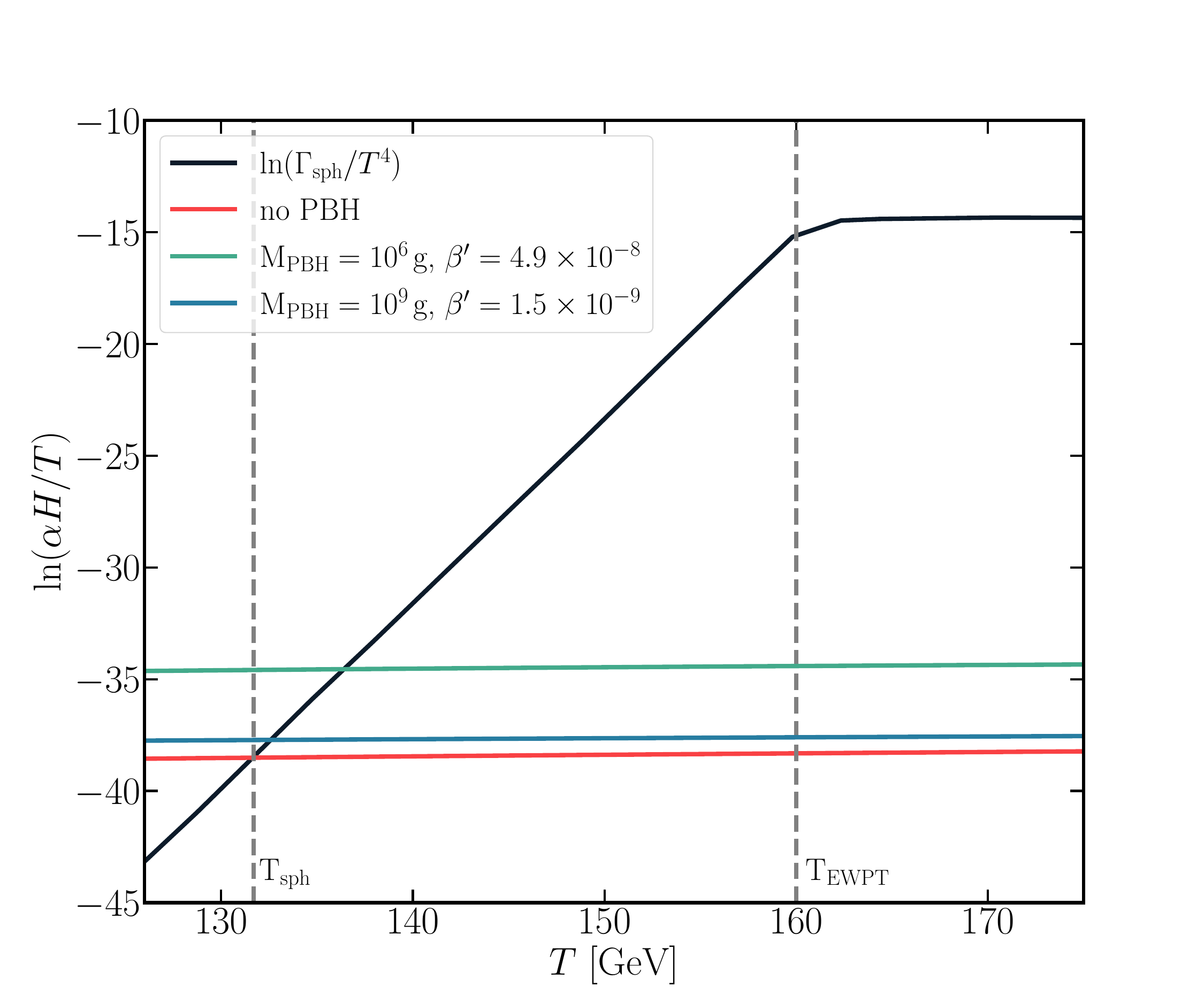}
        \caption{\label{fig:sphaleron}The rate of sphaleron process (black line) as a function of the temperature. The colored lines show the Hubble rate for different scenarios with and without the presence of PBHs. The crossing between $\Gamma_{\rm sph}$ and $H$ defines the temperature $T_{\rm sph}$ at which the sphaleron processes freeze-out. We find $T_{\rm sph} < T_{\rm EWPT}$ in the whole parameter space analyzed.}
\end{figure}

In our scenario, the sphaleron processes go out of equilibrium during a matter-dominated epoch, but always after the electroweak phase transition. The temperature at which sphaleron processes freeze-out is defined by the equation~\cite{DOnofrio:2014rug}
\begin{equation}
    \Gamma_{\rm sph}(T_{\rm sph}) / T_{\rm sph}^3 = \alpha\, H(T_{\rm sph})\,,\label{eq:tsph}
\end{equation}
where $\Gamma_{\rm sph}$ is the rate of the sphaleron processes and $\alpha \approx 0.1015$. In the standard cosmological evolution without PBHs, the solution for $T_{\rm sph}$ is 131~GeV which is smaller than $T_{\rm EWPT} \approx 160~{\rm GeV}$. In the presence of PBHs which drive a matter-dominated epoch with a higher Hubble rate, the temperature $T_{\rm sph}$ is in general higher. However, we have checked that in the considered ranges of PBH mass and abundance, $T_{\rm sph}$ is always smaller than $T_{\rm EWPT}$, which is assumed to be not significantly affected by PBHs.\footnote{Note that in the case of $T_{\rm sph}>T_{\rm EWPT}$, the efficiency factor would differ by less than 10\% which is typically negligible compared to the modification induced by the large entropy injection from PBH evaporation.} This ensures the consistency of our calculations which take $\eta_{\rm sph}=12/37$. In Fig.~\ref{fig:sphaleron} we show the sphaleron rate (black line) as a function of the temperature compared to the Hubble rate obtained without PBHs (red line) and in the presence of PBHs with masses $10^6$ (green line) and  $10^9$ (blue line) in grams, taking their abundance to be the maximal allowed value according to the current constraints.

\begin{figure}[t!]
    \centering
        \includegraphics[width =\linewidth]{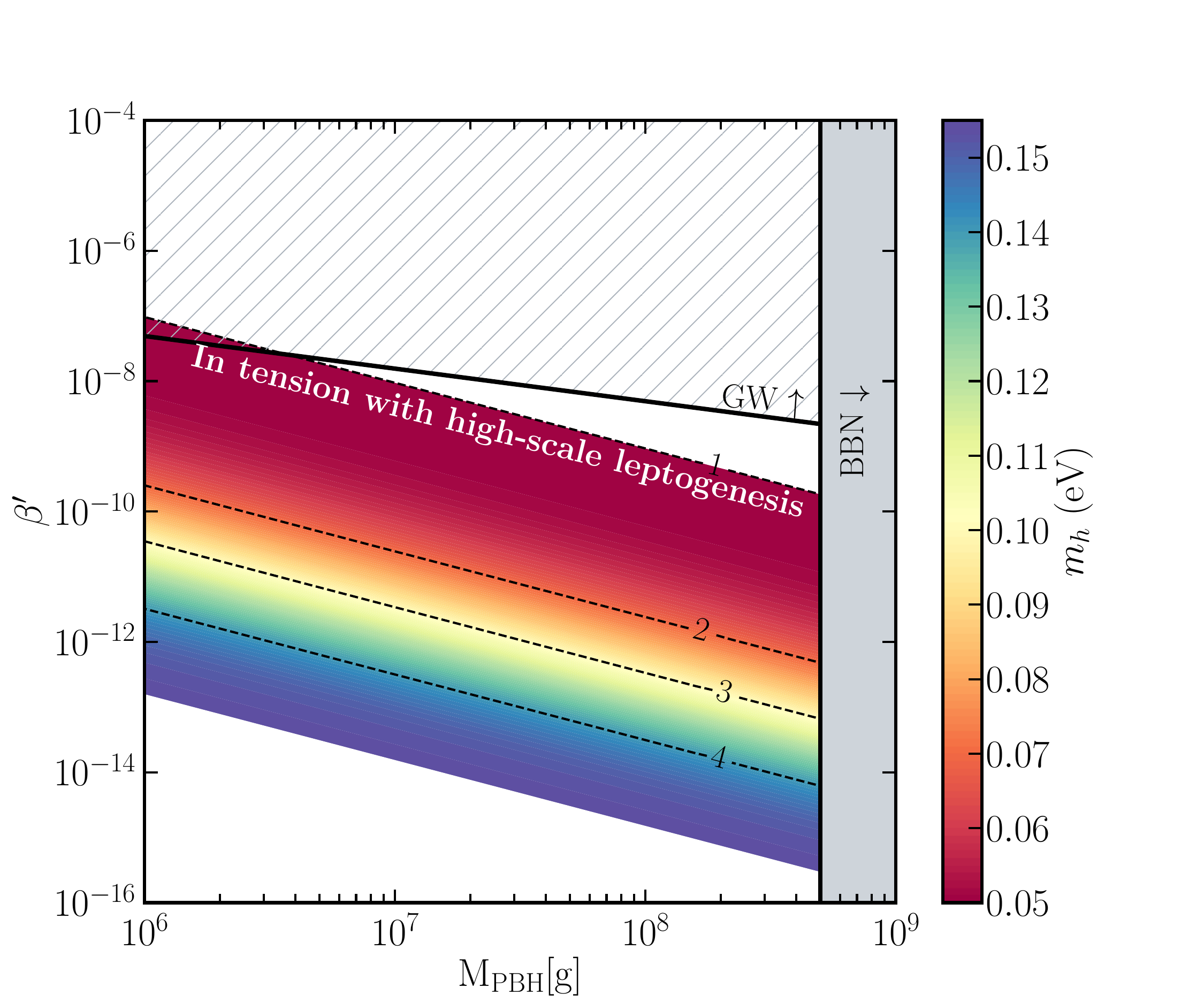}
        \caption{\label{fig:constraints}Mutual exclusion limits between PBHs and the minimal high-scale leptogenesis in the $\MPBH$-$\beta^\prime$ plane. The different colors correspond to the upper bounds obtained assuming different masses $m_h$ for the heaviest active neutrino, while the other neutrino parameters are fixed to maximise the baryon asymmetry. The numbered dashed lines refer to the benchmark cases reported in Fig.~\ref{fig:Ybm}. The hatched region is excluded according to the constraints on the GW energy density from BBN observations~\cite{Domenech:2020ssp}. The shaded grey region highlights PBHs whose evaporation occurs during BBN.}
\end{figure}

The main results of our analysis are displayed in Fig.~\ref{fig:constraints} where we report the mutual exclusion limits between PBHs and high-scale leptogenesis projected in the PBH parameter space. In particular, for each active neutrino mass $m_h$ we derive an upper bound in the PBH parameter space further fixing the right-handed neutrino mass $M_1$ to maximise the baryon asymmetry. In the leptogenesis parameter space, this corresponds to considering the values along the solid black line in Fig.~\ref{fig:Ybm}. Indeed, the four numbered dashed lines refers to the four benchmark cases in Fig.~\ref{fig:Ybm}. The maximum allowed asymmetry in the minimal high-scale leptogenesis decreases as $m_h$ increases, thus placing stronger constraints on $\beta^\prime$. The strongest upper bound on $\beta^\prime$ (darkest violet line) corresponds to the highest possible value for $m_h$ so that the observed baryon asymmetry is achieved. Conversely, the conservative constraint (darkest red line) is set for the lowest possible value for $m_h\approx 0.05~{\rm eV}$ according to neutrino oscillation data. We find that such a conservative bound is stronger than the one imposed by the limit on the energy density of gravitation waves (GWs) at the time of BBN~\cite{Domenech:2020ssp, Papanikolaou:2020qtd} (hatched area in the plot). In the figure, we also highlight the shaded grey area where PBHs would evaporate and dilute the baryon asymmetry during BBN.

\section{Conclusions}\label{sec:concl}

We investigate the impact of the non-standard cosmology driven by the presence and the evaporation of light primordial black holes on the production of the baryon asymmetry of the Universe through high-scale leptogenesis. Crucially, the evaporation of PBHs is associated with (a sudden) injection of entropy and reheating of the Universe. When this occurs after the sphaleron freeze-out, the baryon asymmetry is diluted. In this framework, we firstly explore the four-dimensional parameter space of high-scale leptogenesis in its minimal version demonstrating the existence of a finite, maximum achievable asymmetry $\YB^{\rm max}$. We then point out that, if the entropy injection from primordial black holes' evaporation is large enough, even $\YB^{\rm max}$ does not reproduce the observed asymmetry of the Universe. This effectively rules out high-scale leptogenesis as the baryogenesis mechanism. In this fashion, we derive mutual exclusion limits in the PBH and leptogenesis parameter spaces, delineating the regions which are incompatible. For an entropy increase greater than around four orders of magnitude, the minimal model for high-scale leptogenesis is insufficient to account for the baryon asymmetry of the Universe. This corresponds to a PBH population with masses from $10^6$ to $10^9$ grams with an initial abundance $\beta^\prime \gtrsim 10^{-9}$. Interestingly, we point out the existence of a relation between the heaviest active neutrino mass $m_h$ and the conservative bound in the PBH parameter space which results to be stronger than existing constraints.

\section*{Acknowledgments}

We thank Ofelia Pisanti, Jessica Turner, and Yuber F. Perez-Gonzalez for the useful discussion. This work was partially supported by the research grant number 2017W4HA7S ``NAT-NET: Neutrino and Astroparticle Theory Network'' under the program PRIN 2017 funded by the Italian Ministero dell'Universit\`a e della Ricerca (MUR). We also acknowledge the support by the research project TAsP (Theoretical Astroparticle Physics) funded by the Istituto Nazionale di Fisica Nucleare (INFN).

\bibliographystyle{unsrt}
\bibliography{Bibliography}

\end{document}